\documentclass{article}

\usepackage{amssymb}
\usepackage[english]{babel}
\usepackage{bm}
\usepackage{booktabs}
\usepackage{calc}
\usepackage[font=footnotesize,skip=5pt]{caption}
\usepackage{caption}
\usepackage{chemformula}
\usepackage{color}
\usepackage{commath}
\usepackage[shortlabels]{enumitem}
\usepackage[T1]{fontenc}
\usepackage{float}
\usepackage{gensymb}
\usepackage{geometry}
\usepackage{graphicx}
\usepackage{hanging}
\usepackage{hyperref}
\usepackage{lineno}
\usepackage{lipsum}
\usepackage{mathrsfs}
\usepackage{mathtools}
\usepackage{multicol}
\usepackage{multirow}
\usepackage{soul}
\usepackage{textcomp}
\usepackage[small]{titlesec}
\usepackage{url}
\usepackage{upgreek}
\usepackage{wrapfig}
\usepackage{setspace}

\usepackage[super,numbers,compress]{natbib}

\definecolor{newred}{RGB}{180,20,5}

\definecolor{newlight green}{RGB}{1,129,30}

\definecolor{newblue}{RGB}{40,100,250}

\begin{document}
\twocolumn[
\vspace*{.2cm}
{\centering
{\bf\Large{Direct laser writing of high aspect ratio nanochannels for nanofluidics}}
\vspace{1cm}

Stoffel D.\ Janssens$^{1,*}$, Meissha Ayu Ardini$^1$, David V\'azquez-Cort\'es$^1$, Cathal Cassidy$^2$, and Eliot Fried$^{1,*}$
\vskip 0.5cm
$^1$Mechanics and Materials Unit, Okinawa Institute of Science and Technology Graduate University, 1919-1 Tancha, Onna-son, Kunigami-gun, Okinawa, Japan 904-0495\\
$^2$Science and Technology Group, Okinawa Institute of Science and Technology Graduate University, 1919-1 Tancha, Onna-son, Kunigami-gun, Okinawa, Japan 904-0495\\

\vskip 0.5cm
$^*$Corresponding authors: Stoffel D.\ Janssens (stoffel.janssens@oist.jp), Eliot Fried (eliot.fried@oist.jp)

}
\vskip 0.5cm
\noindent
Nanochannels with high width-to-height aspect ratios are desirable for many applications, particularly those requiring optical access, but remain challenging to fabricate.
In this work, the direct laser writing of such channels between diamond films and glass substrates is introduced.
As previously reported, laser light can transform a portion of diamond film into a nanostrip.
The strip induces delamination of the surrounding film, causing the formation of two nanochannels with triangular cross-sections. 
Here, it is demonstrated that nanochannels with rectangular cross-sections and width-to-height aspect ratios exceeding fifty can form between pairs of nanostrips.
With atomic force microscopy, the maximum strip spacing that produces these nanochannels is investigated, and it is demonstrated that the reflectance of the channels can be measured by microspectrophotometry. 
The microstructure of the nanochannels, including nanostrips, and processes that occur during laser writing are inferred from transmission electron microscopy and electron energy loss spectroscopy.
By fabricating a nanofluidic device and using microspectrophotometry, it is found that the nanochannels fill with water through capillary action, are resistant to clogging, and are mechanically stable against water filling.
A versatile platform for producing high-aspect-ratio nanochannels that are optically accessible and fluidically functional is presented, thereby expanding opportunities for advanced applications.
\vskip 1cm
]

\section{Introduction}
\label{sec:introduction}

Structures with one or more dimensions below 100~nm can display scale-dependent physical behavior significantly different from that predicted by classical bulk descriptions. This reflects the increasing influence of surface, quantum, and confinement effects at the nanoscale. These effects open opportunities for novel applications and technological progress. However, realizing these opportunities requires practical methods for nanostructure fabrication \cite{Schmool2021}.

Laser writing is increasingly applied for the fabrication of complex three-dimensional structures.
By two-photon lithography, such structures can be written down to the submicron scale \cite{Liang2026}. 
Additionally, the possibility exists that transformed bulk can be etched selectively to fabricate elaborate structures \cite{Haward2023}.
Examples of structures that are made by laser writing are
electrically conductive tracks in bulk diamond \cite{Salter2024},
waveguides \cite{Guo2025},
strain sensors \cite{Wang_b_2025},
photonic processors \cite{Skryabin2023},
electrodes \cite{Zhang2026},
polished surfaces \cite{Katamune2023},
and nanochannels \cite{Janssens2023}.

Research on the effects of nanochannels on fluid and particle flow has brought notable progress in 
water desalination and filtration \cite{Tongxi2025}, 
biomedical applications \cite{Wu2025}, 
and energy harvesting \cite{Suman2025}.
These advances arise because the characteristic length of solid--fluid interactions is comparable to the dimensions of nanochannels and because of confinement effects \cite{Shuvo2025}. 
Notably, confinement can cause DNA molecules to elongate \cite{Yi2025}.

Nanochannels, including nanopores and nanoslits, are often fabricated through techniques such as electron beam lithography, focused ion beam milling, reactive ion etching, thin-film evaporation, photolithography, and thermal bonding \cite{Chen2021}. 
These conventional methods can produce functional devices, but they remain complex, expensive, and time-consuming.
A viable alternative to these techniques can be laser writing \cite{Sima2021}.
It is demonstrated that nanochannels can be made by laser writing slits in graphene flakes that subsequently are stacked between two unprocessed flakes \cite{Cui2024}. 
On the sub-micron scale (between 100~nm and 1~$\upmu$m), laser writing followed by chemical etching can produce channels in fused silica \cite{Barbato2024} and silicon \cite{Sabet2024}.
Laser writing of sub-micron nanochannels that are formed by the delamination of thin silica films is also demonstrated \cite{McDonald2006,Bakhtiari2024}.
Blind holes with diameters of less than 100~nm and lengths of less than $200~\upmu$m can be fabricated in fused silica using a laser-writing technique that relies on material ejection \cite{Lu2022,Wang2025}.
In thin nickel films, the formation of nanocavities by spallation is reported \cite{Temnov2020}. 
Such cavities can be applicable to magnetism-related research, but are not optically accessible.
\begin{figure*}[h]
\centering
\includegraphics[scale=1]{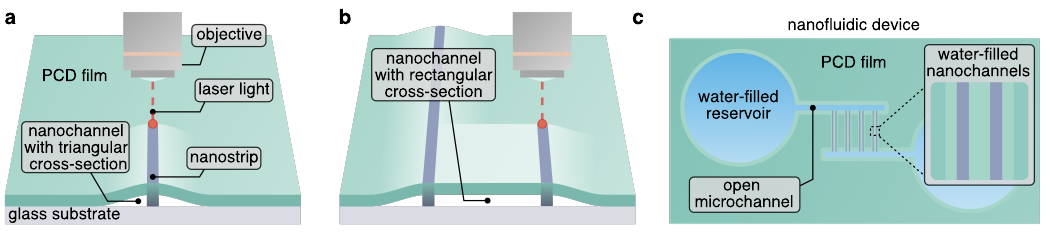}
\caption{{\bf Introduction.} {\bf a}~Schematic illustrating the direct formation of nanochannels between a polycrystalline diamond (PCD) film and a glass substrate by laser writing \cite{Janssens2023}. Laser light locally transforms a portion of the sample into a nanostrip that induces delamination of the surrounding film. The strip consists of non-diamond carbon that supports the delaminated film and originates from diamond through an allotropic transformation. The nanochannels have cross sections that resemble triangular slits and are several micrometers wide. {\bf b}~Schematic of the laser writing process that we investigate in this work. By writing two nanostrips parallel to each other, we demonstrate the formation of nanochannels with rectangular cross-sections and high aspect ratios. {\bf c} Schematic of the nanofluidic device that we use to demonstrate that the channels can fill with water by capillary action.}
\label{fig:schematic}
\end{figure*}

Recently, we introduced the direct formation of nanochannels between thin polycrystalline diamond (PCD) films and glass substrates by laser writing \cite{Janssens2023}. 
This laser writing procedure is schematically depicted in Figure~\ref{fig:schematic}a.
During laser writing, a portion of the material is converted into a nanostrip that occupies a larger volume than the original unmodified region.
As a consequence of this expansion, the film surrounding the nanostrip delaminates and forms nanochannels with cross-sections that resemble triangular slits. 
The nanostrips contain non-diamond carbon that is formed from diamond by an allotropic transformation.
We found that the non-diamond carbon is essential in supporting the nanochannels and that nanostrip formation is initiated in the PCD film close to the film--substrate interface. 
We provided evidence that the latter finding is caused by the defects in that vicinity. 
These defects are more likely to absorb light than pristine diamond and substrate glass.

In the present paper, we describe the application of our femtosecond laser writing and delamination-based approach to the fabrication of rectangular nanochannels between pairs of laser-written nanostrips, together with their microstructural characterization, optical probing, nanofluidic integration, and stability.
Instead of generating paired slit-like channels flanking a single nanostrip, rectangular nanochannels are fabricated between two parallel laser-written nanostrips, as schematically depicted in Figure~\ref{fig:schematic}b, yielding width-to-height aspect ratios exceeding 50 and channel geometries compatible with nanofluidic transport and optical interrogation.
We determine a practical design rule linking nanostrip spacing to the resulting channel height and use it to delineate the processing window for stable rectangular channel formation. 
Through combined transmission electron microscopy and electron-energy-loss spectroscopy, the laser-modified carbon phase is characterized in detail, and the non-diamond carbon responsible for mechanical support is found to be predominantly amorphous, including a distinct layer adjacent to the glass interface. 
We apply microspectrophotometry as an optical probe to these rectangular nanochannels, thereby obtaining spatially resolved reflectance measurements that are not accessible for the slit-like nanochannels formed alongside single nanostrips described in our earlier work. 
These channels are integrated into nanofluidic devices, and capillary water filling is corroborated through reflectance measurements supported by optical simulations. A schematic of that device is depicted in Figure~\ref{fig:schematic}c.
Finally, the stability of a channel is tested against clogging and repeated water filling and draining.

PCD films deposited on glass substrates with chemical vapor deposition (CVD) serve as an optically transparent platform for device fabrication \cite{Janssens2019}.
Both PCD and glass are manufactured from abundant material resources and can be produced in large quantities, which makes the platform commercially viable.
Because the nucleation rate of diamond is too low for practical film growth, substrates are seeded with nanodiamonds \cite{Giussani2022,Vazquez2024}.

Similar to single-crystal diamond \cite{Zhao2025}, PCD can have outstanding properties.
For example, PCD exhibits a high Young’s modulus \cite{Williams2010}, which can make it beneficial for MEMS applications \cite{Huber2026}.
Additionally, PCD can have a high thermal conductivity, which is being investigated for the thermal management of semiconductor devices \cite{Misono2025,Moriyama2025,Ning2025}. 
PCD can also be doped to make it electronically conductive \cite{Janssens2011,Kato2016}. 
Boron-doped diamond has excellent electrochemical properties and offers opportunities in the field of supercapacitors \cite{Liao2024,SumanShradha2025}. 
Furthermore, the wide range of applications of diamond in chemistry, biology, and quantum sensing \cite{Nianjun2025} suggests that incorporating diamond into nanofluidic channels has promising prospects for future applications.

\begin{figure*}
\centering
\includegraphics[scale=1]{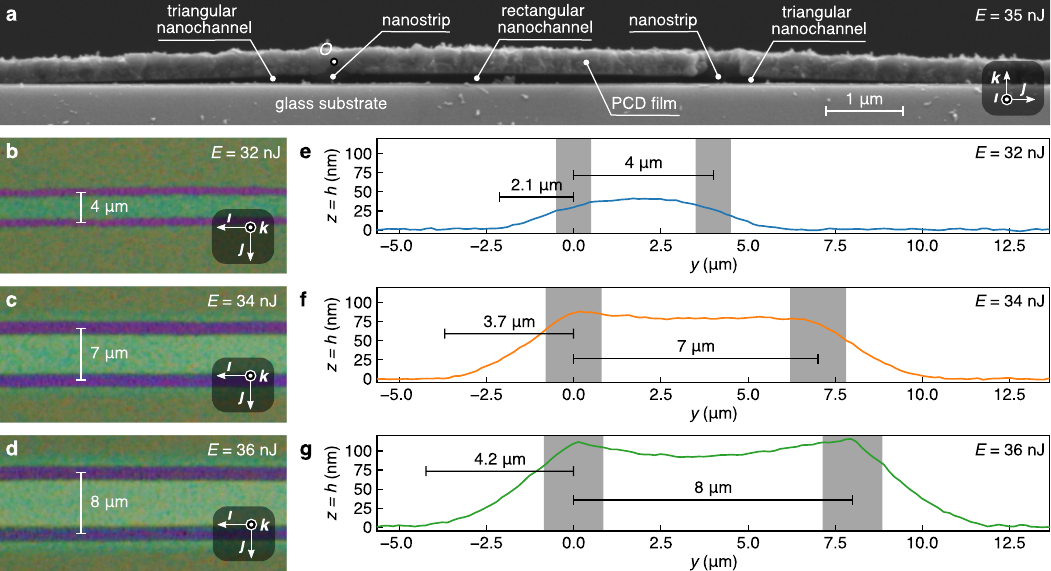}
\caption{{\bf Nanochannel shape.} {\bf a}~Scanning electron microscopy image of a rectangular nanochannel formed between two nanostrips. Each nanostrip is flanked by a triangular nanochannel. The orthonormal basis of a rectangular Cartesian coordinate system with axes $x$, $y$, and $z$ and origin $O$ is formed by vectors $\bm{\imath}$, $\bm{\jmath}$, and $\bm{k}$. Quantity $E$ denotes laser pulse energy, and $O$ is located at the film--air interface before laser writing. {\bf b}--{\bf d}~Reflected light microscopy image of nanochannels. {\bf e}--{\bf g}~Profile height $h$ of the structures in ({\bf b}--{\bf d}), respectively, obtained by averaging atomic force microscopy (AFM) data over $x$. The gray regions indicate the $y$-positions of the nanostrips, and a scale bar located at $y \leq 0$ coincides with the width of a corresponding triangular nanochannel.}
\label{fig:shape}
\end{figure*}
\section{Results and discussion}
\subsection{Nanochannel shape}
\label{subsec:shape}
To produce a structure that resembles the cross-section of a rectangular nanochannel formed between a pair of nanostrips, each flanked by a triangular channel, we cleave a sample into two similar parts and perform laser writing over a cleaved edge.
A scanning electron microscopy image of the obtained structure is provided in Figure~\ref{fig:shape}a, where $E$ represents laser pulse energy. Vectors $\bm{\imath}$, $\bm{\jmath}$, and $\bm{k}$ form the orthonormal basis of a rectangular Cartesian coordinate system with axes $x$, $y$, and $z$ and origin $O$.
Because scanning electron microscopy provides limited information on the nanostrip microstructure, a combined transmission electron microscopy (TEM) and electron energy loss spectroscopy (EELS) study is presented in \secref{subsec:TEM_EELS}.

In Figure~\ref{fig:shape}b--d, reflected light microscope images of nanochannels fabricated on the same sample and for several values $E$ are depicted.
The nanostrips are purple, and the other portions of the sample, including the nanochannels, are green.

The profile height $h$ of the structures is plotted as a function of $y$ in Figure~\ref{fig:shape}e--g. Height $h$ is measured relative to $O$, which is always positioned at the film--air interface before laser writing, and is obtained by averaging atomic force microscopy (AFM) data over $x$.
From the plots in this figure, we infer that the height of the channels increases as $E$ increases. 
Consequently, we deduce that the brightness of the nanochannels in Figure~\ref{fig:shape}b--d increases as their height increases.
From these observations, we infer that the reflectance of the channels depends on channel height.
The results of a microspectrophotometry study conducted to investigate this are presented in \secref{subsec:Reflectance}.

From Figure~\ref{fig:shape}b--d, we find that the nanostrips increase in width as $E$ increases. 
This is because the spot at which sample material is transformed by laser light increases in diameter as $E$ increases \cite{Liu1982}.
From Figure~\ref{fig:shape}e--g, we find that at a nanostrip the curvature of a surface profile increases as $E$ increases. 
This can be caused by a decreased mechanical stability of the PCD films, which makes the films more pliable at those locations.
A decrease in mechanical stability is expected when diamond is transformed into non-diamond carbon, which typically has a lower Young's modulus than that of diamond \cite{Klein1993}.
This transformation is more pronounced as $E$ increases, which explains the observed trend.

We find that the profile heights of the nanostrips fabricated by the same settings of our laser writing system vary significantly.
We attribute these fluctuations to operation in the low-power regime, where the laser exhibits reduced stability compared to the intermediate-power regime. 
In future work, we plan to mitigate this issue by operating in the more stable intermediate-power range, achieved through optical attenuation of the laser beam.

\begin{figure}[h!]
\centering
\includegraphics[scale=1]{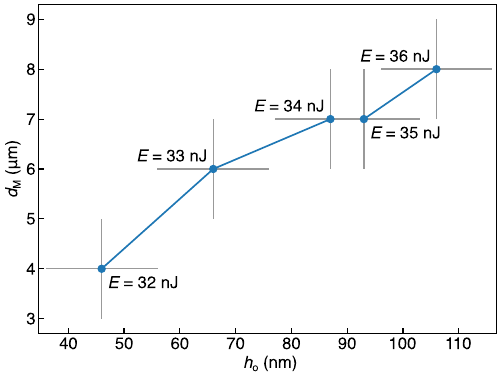}
\caption{{\bf Maximum nanostrip spacing.} Maximum spacing $d_{\text{M}}$ between the center lines of two adjacent nanostrips that still allows the formation of rectangular nanochannels, plotted as a function of the overall profile height $h_{\text{o}}$ of the resulting nanochannel. The height $h_{\text{o}}$ is determined by averaging the local channel height $h$ across the region between the center lines of the two nanostrips. Corresponding values of $E$ associated with each $h_{\text{o}}$ value are indicated next to the respective data points. The uncertainty in $d_{\text{M}}$ is the step size ($1~\upmu\text{m}$) used to measure that quantity, while the uncertainty in $h_{\text{o}}$ is the root mean square roughness of the PCD film (measured by AFM), which is significantly larger than other sources of uncertainties.}
\label{fig:w_h}
\end{figure}
From the surface profiles in Figure~\ref{fig:shape}e--g, we observe that the width of triangular channels increases as $E$ increases.
As a result, the maximum spacing $d_\text{M}$ between the center lines of a pair of strips that still permits the formation of rectangular channels also increases.
We also find that the overall profile height $h_\text{o}$, calculated by averaging $h$ across the $y$-axis from the center of one nanostrip to the center of the adjacent nanostrip, increases as $E$ increases.
From these observations, we infer that $d_\text{M}$ scales with $h_\text{o}$.
This trend is confirmed by plotting $d_\text{M}$ against $h_\text{o}$ in Figure~\ref{fig:w_h}.
By taking the width and the height of a rectangular channel as $d_\text{M}$ and $h_\text{o}$, respectively, we find that the width-to-height ratio of our channels significantly exceeds fifty.

\begin{figure*}[h]
\centering
\includegraphics[scale=1]{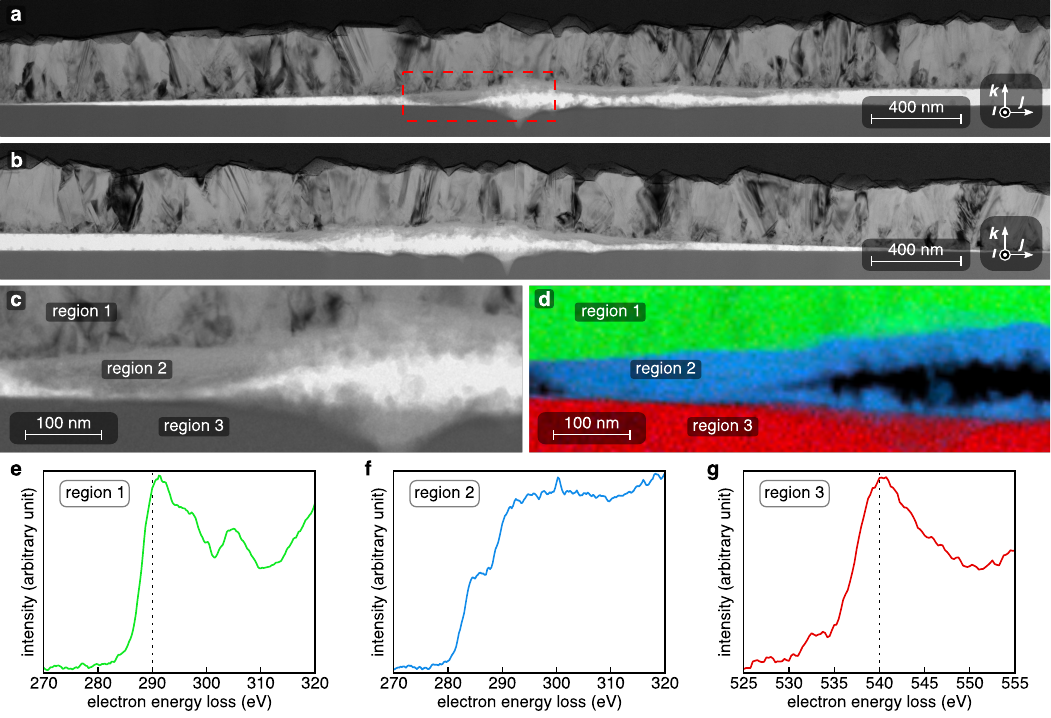}
\caption{{\bf Microstructure.} {\bf a}, {\bf b} Panoramic transmission electron microscopy (TEM) images of a representative nanochannel fabricated with $E = 35$~nJ. {\bf c} Enlarged view of the area within the dashed rectangle in ({\bf a}). {\bf d} Electron energy loss spectroscopy (EELS) map of the area in ({\bf c}), obtained by investigating K edges. {\bf e}--{\bf g} Representative EELS spectra of the regions in ({\bf d}). By comparing these spectra with those in the literature \cite{Chu2006,Ayoola2020}, we infer that regions 1, 2, and 3 contain diamond, amorphous carbon, and oxygen (substrate), respectively. The peak around 290~eV in ({\bf e}) is a signature of diamond, and the absence of peaks in ({\bf f}) corresponds to amorphous carbon. The peak around 540 eV in ({\bf g}) is characteristic of oxygen.}
\label{fig:microstructure}
\end{figure*}

To clarify the origin of the rectangular channel cross-sections formed between pairs of laser-written nanostrips, a qualitative mechanistic interpretation can be introduced. 
We interpret the formation of rectangular nanochannels between two parallel nanostrips as arising from the interaction of two delamination processes initiated adjacent to each strip. 
In the vicinity of each nanostrip, femtosecond laser writing produces a locally transformed carbon region accompanied by residual strain and reduced interfacial adhesion relative to the surrounding film. 
In the single-strip configuration described in our previous work, relaxation of this mismatch proceeds through delamination on both sides of the strip, and the resulting cavity typically adopts a wedge-like cross-section, consistent with a peel-type opening that reduces elastic energy at limited interfacial area cost. 
In the two-strip configuration considered here, delamination is initiated adjacent to each strip; when the spacing between strips is sufficiently small, the two delaminated regions approach one another before either can evolve into an independent wedge-like cavity. 
Under these conditions, a configuration in which the detached film segment spanning the two strips remains attached at the strip locations and separates from the substrate primarily between them is inferred to reduce bending energy relative to two isolated wedges. 
The resulting cross-section consists of a central region with mild curvature bounded by localized regions of relatively high curvature near the strips, giving rise to a roof--floor separation that is approximately rectangular. 
On the basis of this interpretation, we account for the existence of a maximum strip spacing for rectangular-channel formation: when the spacing exceeds a critical value, the delaminated regions no longer overlap, and relaxation proceeds instead through two largely independent wedge-like cavities adjacent to each strip.

\subsection{Microstructure}
\label{subsec:TEM_EELS}
In our previous work \cite{Janssens2023}, we found through Raman spectroscopy that laser writing causes the transformation of diamond into non-diamond carbon.
By scanning electron microscopy, we found that this transformation occurs near the substrate--film interface.
Because the portion of the PCD film close to that interface contains more defects, we hypothesized that laser light is absorbed there the most, leading to this transformation.
By selectively removing non-diamond carbon at 773~K in air, we observed a significant decrease in channel height, which led us to infer that the non-diamond carbon supports the PCD film.
At even higher temperatures, we removed the entire PCD film and, using AFM, we found that the glass substrate was also affected by laser writing.
In this section, we present a combined TEM and EELS study to further elucidate the microstructure of our nanochannels and nanostrips.

\begin{figure*}[h]
\centering
\includegraphics[scale=1]{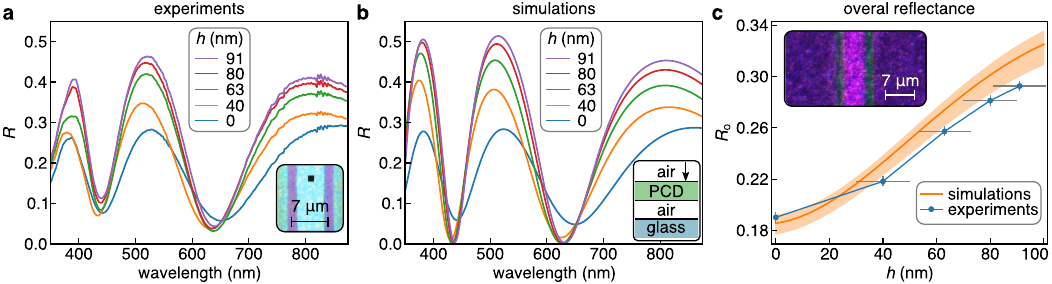}
\caption{{\bf Reflectance versus channel profile height.} {\bf a}~Specular reflectance $R$ measured by microspectrophotometry for a $286\pm10$~nm thick PCD film on glass ($h=0$) and for nanochannels ($h>0$). The inset is a reflected light microscopy image of a rectangular nanochannel above which a mirror is placed. The mirror is the black square and is used to collect reflected light for the measurement of $R$, which is performed at the center of a channel. {\bf b}~Simulated values of $R$ for channel heights equal to $h$. The inset is a schematic of the model used for the calculations. The arrow indicates the propagation direction of the incident light beam. {\bf c}~Overall reflectance $R_\text{o}$ of $R$. The orange band represents the deviation in film thickness, which corresponds to the root mean square surface roughness of the film. For the experimental data, the uncertainty in $h$ is the surface roughness, and the uncertainty in $R_\text{o}$ is a standard deviation calculated from three independent measurements. The inset is a dark-field microscopy image of a rectangular nanochannel.}
\label{fig:reflectance}
\end{figure*}

The TEM images in Figure~\ref{fig:microstructure}a,~b show the nanostrips of a representative channel fabricated with $E = 35$~nJ. 
These images were produced by milling a lamella from the sample and stitching together multiple individual TEM images acquired from this lamella.
In Figure~\ref{fig:microstructure}a, the center of the rectangular nanochannel lies to the right of the nanostrip; in Figure~\ref{fig:microstructure}b, it lies to the left.
Figure~\ref{fig:microstructure}c is an enlarged view of the area within the dashed rectangle in Figure~\ref{fig:microstructure}a.

Figure~\ref{fig:microstructure}d is an EELS map, obtained by investigating K edges, that corresponds with the area of Figure~\ref{fig:microstructure}c. 
Figure~\ref{fig:microstructure}e--g are characteristic EELS spectra found in regions 1, 2, and 3, respectively. 
From the literature \cite{Chu2006,Ayoola2020}, we find that the spectra in Figure~\ref{fig:microstructure}e--g correspond to those typically found for diamond, amorphous carbon, and oxygen (glass substrate), respectively. 
In Figure~\ref{fig:microstructure}e, the peak at approximately 290~eV is characteristic of diamond. 
The absence of peaks in Figure~\ref{fig:microstructure}f is a clear signature for amorphous carbon, and the peak around 540~eV in Figure~\ref{fig:microstructure}g corresponds to that of oxygen.
We conclude that the non-diamond carbon formed during laser writing is predominantly amorphous carbon.
Other studies on the laser-induced transformation of diamond into amorphous carbon are summarized in a review \cite{Ali2021}.
The amorphous carbon forms a layer on the PCD film and a thinner layer on the glass substrate, and in Figure~\ref{fig:microstructure}c, the layer is jammed between the PCD film and the glass substrate. 
Based on this observation, we infer that the amorphous carbon layer supports the delaminated portions of the PCD film.

Figure~\ref{fig:microstructure}a and Figure~\ref{fig:microstructure}b are not exactly mirror images.
This can be explained by the fact that the centers of the laser spots are not precisely aligned with the location where the lamella is extracted from the sample.
Slight variations in $E$ and PCD film quality might also play a role.
The substrate is slightly affected by the laser writing process, and we find the presence of carbon-containing nanoparticles.

\begin{figure*}
\centering
\includegraphics[scale=1]{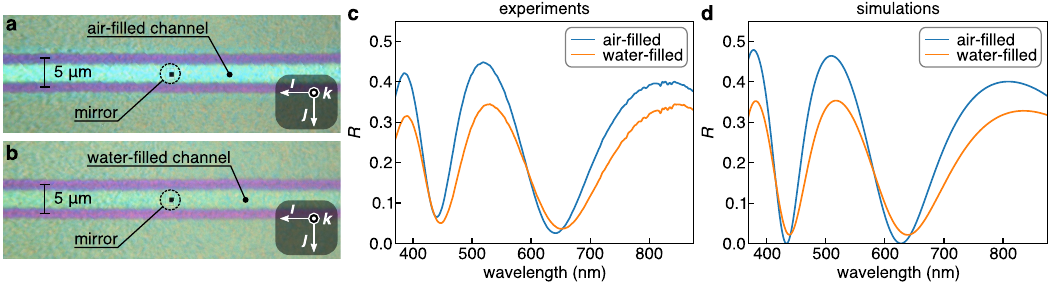}
\caption{{\bf Reflectance of water-filled channel.} {\bf a}~Reflected light microscopy image of a rectangular nanochannel in a nanofluidic device. A mirror is placed above the channel for measuring $R$. The nanostrips are written with an energy of $E = 35$~nJ, and the spacing between the center lines of the strips is $5~\upmu\text{m}$. The PCD film is $286\pm10$~nm thick, and below the mirror, $h = 65 \pm 10$~nm. The channel is connected at both ends to open microchannels that lead to air-filled reservoirs. {\bf b} Portion of the same channel as in ({\bf a}) after placing water in the reservoirs that causes the channel to fill with water through capillary action. {\bf c} Reflectance $R$ of the nanochannel in air-filled and water-filled states. {\bf d} Simulated values of $R$ for the nanochannel.}
\label{fig:reflectance_device}
\end{figure*}

\subsection{Reflectance}
\label{subsec:Reflectance}
Because our rectangular nanochannels are relatively flat and wide, we can apply microspectrophotometry to investigate the effect of profile height on the reflectance $R$ of the nanochannels.

\subsubsection{Reflectance versus channel profile height}
\label{sec:r_v_c_h}
In Figure~\ref{fig:reflectance}a, $R$ is provided for a $286\pm10$~nm thick PCD film on glass ($h = 0$) and for several nanochannels ($h > 0$). 
For the nanochannels, $R$ is measured in the center of a channel.
This is clarified by the inset, in which a reflected light microscopy image of a nanochannel is displayed. 
The black square is a mirror that is placed in the optical path and is used to locally collect light for measuring $R$. 
We find that for a large portion of wavelengths, $R$ increases as $h$ increases. 
For example, at a wavelength of 525~nm, $R$ approximately increases from $0.28$ ($h=0$) to $0.46$ ($h = 91$~nm), so that $\Delta R = 0.18$.

To support the results in Figure~\ref{fig:reflectance}a, the simulated values of $R$ are computed, which is done using the transfer matrix method \cite{Fontanot2023}.
The results of computations where $h$ is taken as channel height are depicted in Figure~\ref{fig:reflectance}b.
The inset is a schematic of the model that we used for the computations. 
The arrow that is perpendicular to the PCD film indicates the propagation direction of the incident beam. 
As $h$ increases, we find that the experimental and simulated values of $R$ change in a similar fashion. 
For 525~nm wavelength, we find that the simulated value of $R$ increases from $0.28$ ($h=0$) to $0.51$ ($h = 91$~nm), so that $\Delta R = 0.23$.
This difference is larger than that for the experiments ($\Delta R = 0.18$).
Reasons for this mismatch are provided in the discussion below.

To further analyze $R$, we investigate the behavior of overall reflectance $R_\text{o}$, which is $R$ averaged over wavelength at a fixed value of $h$. 
Experimental and simulated values for $R_\text{o}$ are provided in Figure~\ref{fig:reflectance}c. 
With the orange band, we account for the standard deviation in film thickness due to surface roughness.
We find that the experimental $R_\text{o}$ values of the nanochannels are systematically below those obtained through simulations.
This discrepancy can have several causes. 
For example, in our model, we assume that a rectangular channel is flat.
However, this is not the case, as evidenced by the profiles in Figure~\ref{fig:shape}e--g.
Deviations from flatness can affect light propagation, resulting in lower experimental $R_\text{o}$ values compared to those obtained through simulations.
We also assume that the refractive index of the PCD film is homogeneous.
In reality, this is not the case because the defect density in the PCD film decreases in the direction of $\bm{k}$, and such defects can affect $R_\text{o}$.
The films also have a root mean square surface roughness of 10~nm, as measured by AFM, which is not accounted for in our model.
Surface roughness causes light scattering, which results in lower experimental $R_\text{o}$ values compared to those obtained through simulations.
In addition, we found that the channels can contain nanoparticles.
These particles can scatter light and cause a reduction in $R_\text{o}$.
With the inset of Figure~\ref{fig:reflectance}c, which is a dark-field microscopy image, we demonstrate that light is indeed scattered at a nanochannel. 
In these images, areas of the sample where light is scattered appear brighter than those where less light is scattered.

\subsubsection{Reflectance of water-filled channels}
\label{subsec:nanofluidic_device}
To test if the nanochannels fill with water, we fabricated a nanofluidic device in which both ends of the channels are connected to open microchannels that terminate in reservoirs.
A schematic of that device is depicted in Figure~\ref{fig:schematic}c.
The idea is that by filling the reservoirs with water, capillary action will cause water to fill the open microchannels, and subsequently, the nanochannels.
At the center of a nanochannel, $h = 65 \pm 10 $~nm, the spacing between the center lines of nanostrip pairs is $5~\upmu\text{m}$, the length of the nanostrips is $200~\upmu\text{m}$, the nanostrips are produced with $E = 35$~nJ, and the PCD film is $286\pm10$~nm thick.
More fabrication details on the microchannels and reservoirs can be found in \secref{subsec:nanofluidic_dev_fab}.

A reflected light microscopy image of an air-filled nanochannel of our nanofluidic device is shown in Figure~\ref{fig:reflectance_device}a.
After taking this image, we placed water in the reservoirs and, subsequently, took the reflected light microscopy image displayed in Figure~\ref{fig:reflectance_device}b.
We observe that the nanochannel is significantly less bright than before placing water in the reservoirs, from which we infer that the channel is filled with water.

To quantify the change in brightness, we locally measure $R$ at the center of the nanochannel before and after water filling.
The results of our experiments are shown in Figure~\ref{fig:reflectance_device}c.
The mirror used to collect reflected light and measure $R$ is shown in Figure~\ref{fig:reflectance_device}a,~b.
We find that, at most wavelengths, $R$ is greater when the channel is filled with air than when it is filled with water.
To further analyse $R$, we define $\Delta R_\text{o}$ as $R_\text{o}$ before water filling minus $R_\text{o}$ after filling.
As expected from Figure~\ref{fig:reflectance_device}c, we find that $\Delta R_\text{o}$ is positive ($\Delta R_\text{o} = 0.059$).

To support the results in Figure~\ref{fig:reflectance_device}c, we simulate $R$, which is done in a similar fashion as for $R$ in Figure~\ref{fig:reflectance}b.
The results are shown in Figure~\ref{fig:reflectance_device}d.
From these results, we again find that, at most wavelengths, $R$ is greater when the channel is filled with air than when it is filled with water.
This strongly supports that our channels are indeed filled with water.
For our simulated results, $\Delta R_\text{o} = 0.064$, which is a value that is slightly greater than that of our experimental values ($\Delta R_\text{o} = 0.059$).
We refer to the last paragraph of \secref{sec:r_v_c_h} for a discussion on discrepancies in simulated and experimental reflectance values.

\begin{figure}
\centering
\includegraphics[scale=1]{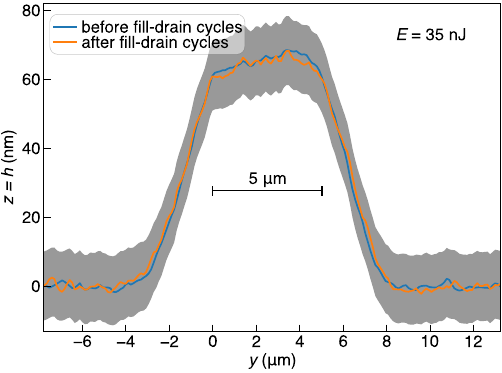}
\caption{{\bf Channel stability.} Profile heights of the nanochannel in Fig.~\ref{fig:reflectance_device}a before and after 101 water fill-drain cycles. During these cycles, the nanofluidic device containing the nanochannel is kept on a 333~K hotplate to accelerate drying-induced nanochannel draining. The grey band represents the deviation in film thickness.}
\label{fig:profile_repeated_filling}
\end{figure}
\subsection{Channel stability}
The stability of the nanochannel depicted in Figure~\ref{fig:reflectance_device}a was tested by repeatedly filling the channel with water and draining it through evaporation.
This procedure was accelerated by placing the device containing the channel on a 333~K hotplate and was conducted under otherwise ambient conditions.
The channel remained operable after 101 cycles, which indicates stability against clogging.
Reflected light microscopy images of the channel in the water- and air-filled (drained) states at each cycle are provided as supplementary information.

Figure~\ref{fig:profile_repeated_filling} shows the profile heights of the channel before and after the hundred cycles.
The grey band accounts for the deviation in film thickness.
The two profiles are practically indistinguishable, which demonstrates that the channels are mechanically stable against repeated water filling at 333~K.

We note that amorphous carbon is stable in air up to at least 573 K \cite{Bannov2020}, while both PCD and the glass substrate remain stable at significantly higher temperatures \cite{Janssens2023}. 
Together, this indicates that the nanochannels can withstand elevated temperatures in air.

\section{Conclusions}
We established a method for fabricating nanochannels with width-to-height aspect ratios exceeding fifty.
Direct laser writing is applied to a polycrystalline diamond (PCD) film grown on a glass substrate.
The procedure induces localized modifications and delamination, resulting in the formation of nanochannels at the PCD--glass interface with cross-sections that closely resemble rectangular slits.

To fabricate a channel with a rectangular cross-section, we transformed a portion of the sample into a nanostrip. 
This process resulted in the formation of two nanochannels with triangular cross-sections, which were formed through the delamination of the PCD film.
By writing a second nanostrip parallel to the first nanostrip, a channel with a rectangular cross-section was formed.
With atomic force microscopy, we characterized the attainable channel cross-sections and found that channel height is adjustable over a wide range.
We also determined the maximum achievable width-to-height aspect ratio as a function of channel height.

The microstructure of the channels was investigated with transmission electron microscopy and electron energy loss spectroscopy.
At a nanostrip, a portion of the PCD film was found to transform into amorphous carbon.
The amorphous carbon, located between the PCD film and the glass substrate, forms a thicker layer on the PCD film than on the glass substrate.
The preferential absorption and transformation of PCD into amorphous carbon close to the film--substrate interface is attributed to the presence of defects, which are more likely to absorb laser light than the more pristine diamond found on the other side of the PCD film and than the substrate glass.

Microspectrophotometry was applied to probe these nanochannels and, to our knowledge, this approach has not been used previously for this purpose.
The use of microspectrophotometry was enabled by the flatness and lateral extent of the channels.
A clear trend in reflectance spectra was observed as a function of channel height, in close agreement with our simulations.
The overall reflectance of the channels increases with channel height, and deviations from simulated spectra were discussed.

A nanofluidic device was fabricated to demonstrate channel functionality. 
The channels were shown to fill with water by capillary action, and the presence of water was detected by a significant change in channel reflectance, accurately predicted by our simulations.
Through repeated filling and draining experiments, we found that the channels remain mechanically stable and resistant to clogging over more than one hundred cycles.

From these results, we infer that direct laser writing in PCD-glass systems provides a versatile route for the cleanroom-free formation of high-aspect-ratio nanochannels suitable for lab-on-a-chip systems and microspectrophotometric studies.

\section{Materials and methods}
\label{sec:Materials_methods}

\subsection{Polycrystalline diamond film synthesis}
In our earlier work, the diamond films were described as nanocrystalline diamond (NCD).
In this work, similar films are used, but we opted for a more general terminology, namely PCD.
For PCD film synthesis, $10 \times 10 \times 0.6$~mm$^3$ Corning Lotus NXT glass substrates, cut from a 0.6~mm-thick plate using a DISCO DAD322 dicing saw, are used.
The substrates are seeded with nanodiamonds \cite{Chang2022}.
Essential details regarding our seeding method are provided in previous work \cite{Janssens2019}.
To ensure a reproducible seed density, seeding is carried out within one hour of substrate cleaning \cite{Pobedinskas2021}.

An SDS6500X microwave plasma-assisted CVD system of Cornes Technologies is used for synthesizing the PCD films.
A mixture of 495~sccm hydrogen gas and 5~sccm methane gas flows into the reactor of the system, while a Kashiyama SDE90X dry pump maintains the deposition pressure constant at 3.3~kPa, and the gas mixture is ignited into a plasma with 2.5~kW of 2.45~GHz microwaves. 
These conditions maintain the substrate temperature below 995~K, which is the annealing point of the glass, and result in a deposition rate of approximately 2~nm/min.
Results for nanochannel writing are obtained using PCD films that are approximately 290~nm thick.

\subsection{Direct femtosecond laser writing}
A LightFab system is used for nanochannel writing.
The system features a 5.2~W Amplitude Systems Satsuma laser that produces 1033~nm wavelength infrared light.
The system also has a galvo scanner head for writing, and has a ZEISS objective with a 10~mm working distance, a numerical aperture of 0.4, and $20 \times$ magnification. 
To ensure that the laser light is focused on the sample surface, we utilized the autofocus function of the system.
The length of the nanochannels was limited by the galvo scanner head to approximately 700~$\upmu\text{m}$.
The laser light is focused to a spot with a radius of approximately 1~$\upmu\text{m}$, with the laser pulse energy $E$ being the only parameter varied.
Laser writing is performed with a pulse frequency of 756~kHz, a pulse duration of 270~fs, and a write speed of 200~mm/s.
As schematically shown in Figure~\ref{fig:schematic}, a sample is irradiated from the PCD film side at an incidence that, on average, is normal.

\subsection{Nanofluidic device fabrication}
\label{subsec:nanofluidic_dev_fab}
To fabricate our nanofluidic device, we follow the same strategy as outlined in the supplementary data of our previous work \cite{Janssens2023}. The fabrication process begins with the synthesis of a 50 nm thick PCD film on a glass substrate. Next, portions of the PCD film are removed using laser writing on locations where microchannels and reservoirs are intended to form. The thickness of the PCD film is then increased to approximately 290 nm. Due to the low nucleation rate of diamond, the areas that are laser-patterned remain uncovered. After this, the uncovered portions of the glass substrate are etched several micrometers deep using a mixture of hydrochloric acid and buffered oxide etchant. During this procedure, the microchannels and reservoirs are formed. Finally, nanochannels are written between two open microchannels. All of these procedures can be carried out without the need for a cleanroom. The channels can be filled with water using a pipette.

\subsection{Lamella fabrication}
Electron-transparent lamellas for TEM and EELS investigations are prepared using a Helios 650 Focused Ion Beam system (ThermoFisher Scientific), employing the standard lift-out methodology with 30 kV Ga ion milling. 
Given the suspended nature of the nanochannel structure, the lamella is left slightly thicker to minimize mechanical distortions of the structure. 
As a result, the specimen is approximately 1.6 mean free paths thick in the region of the highly scattering diamond layer. 
This is thicker than optimal for EELS measurements, but for qualitative mapping, it is sufficient, and is considered a reasonable compromise to minimize damage to the mechanical integrity of the nanochannel. 
As a reference, a thinner specimen is also prepared from a non-channel region, with a thickness of approximately 0.4 mean free paths in the diamond layer.

\subsection{Characterization}
During CVD, the PCD film thickness is monitored using a home-built interferometer. 
Details of how the film thickness is extracted from the interferometer signal are described in previous work \cite{Vazquez2023}. 
After CVD, film thickness is more accurately measured with spectral interferometry using a Hamamatsu Photonics C13027-11 Optical NanoGauge system.

Scanning electron microscopy is done with a JEOL JSM-7900F system.
To avoid charging effects from occurring during scanning electron microscopy, PCD films are coated with approximately 15~nm of Pt-Pd, which is done using a Hitachi MC1000 Ion Sputter Coater system.

AFM measurements are conducted using a Bruker Dimension ICON3 system operating in PeakForce Tapping mode. A ScanAsyst-Air probe (Bruker) with a silicon tip on a silicon nitride lever is used. The lever has a thickness of $650~\text{nm}$, a length of $115~\upmu\text{m}$, a width of $25~\upmu\text{m}$, a nominal spring constant of $0.4~\text{N/m}$, and a resonance frequency of approximately $70~\text{kHz}$. Imaging is performed over a $50~\upmu\text{m} \times 50~\upmu\text{m}$ area, at a resolution of $512 \times 512~\text{pixels}$, and with a scan rate of approximately $1~\text{Hz}$.

Transmission Electron Microscopy data is acquired using a Titan G2 80-300 TEM (ThermoFisher Scientific). 
For this work, the microscope is operated at an accelerating voltage of 300~kV. 
The microscope is equipped with a Schottky XFEG source and a post-specimen spherical aberration corrector for the TEM mode (CEOS GmbH). 
For EELS measurements, the microscope is equipped with a Quantum 966 GIF (Gatan).
The microscope is also equipped with a pre-GIF 4~k Ceta camera (ThermoFisher Scientific) and with a post-GIF 2~k UltrascanXP (Gatan).

Reflectance spectra are obtained through microspectrophotometry using a 2030PV system from CRAIC Technologies.
We obtain $R$ on the PCD film side of a sample using relation
\begin{equation}
R = \frac{I - I_\text{b}}{I_\text{r} - I_\text{b}} R_\text{r},
\label{eq:reflectance}
\end{equation}
where $I$, $I_\text{b}$, $I_\text{r}$ are reflected light intensities measured by a photodetector for a sample, the background, and a reflectance standard, respectively, and where $R_\text{r}$ is the reflectance of the standard. Obtaining the spectra for \secref{sec:r_v_c_h} is carried out using a ZEISS objective with a 0.31~mm working distance and $100 \times$ magnification.
For \secref{subsec:nanofluidic_device}, a Nikon objective is used with a 2~mm working distance and $100 \times$ magnification.
Reference spectra are obtained using an OptoSigma TFAN-25C05-10 aluminum mirror. 
The greater working distance of this objective allows a droplet of distilled water to remain in the reservoir of the nanofluidic device without coming into contact with the objective.
The tabulated reflectance values for that mirror are used to calculate the experimental $R$ values.

Dark-field images are obtained using an Olympus BX53M microscope with a $50 \times$ UIS 2 objective.
The water fill-drain cycles of a nanochannel are observed with the reflected light microscope of a Keyence VK-X150 using a tungsten light source and a Nikon objective with $10 \times$ magnification. For cycle 101, a Nikon objective with a $50 \times$ magnification is used to produce high-resolution images of the channel at the end of the cycles.
Throughout the filling and draining, the device containing the nanochannels is placed on a hotplate (MSA Factory microscope heat stage).

\subsection{Software and simulations}
The analysis of AFM spectra is conducted with Gwyddion software.
The Python programming language is used to process experimental data, to perform simulations, and to plot data. 
Images are processed with GIMP (GNU Image Manipulation Program), and all figures are constructed using Inkscape.

EELS data is acquired and processed using the Digital Micrograph software package (Gatan). 
For acquisition, convergence and collection angles of 10~mrad and 40.5~mrad are employed, respectively. 
A dual-EELS mode is employed, allowing simultaneous acquisition of the zero-loss peak (for energy offset correction) and the core edges of carbon and oxygen. 
A spectrometer dispersion of 0.25~eV per channel is employed. 
Data is acquired using a spectrum image mode, with a low-loss and core-loss spectrum associated with each pixel. 
The spatial dimensions of the spectrum mapping are $170 \times 60$ pixels, with a step size of 5.1 nm. 
After acquisition, all datasets are corrected for slight energy shifts using the simultaneously acquired zero-loss peak. 
Background subtraction is performed using a power law fitting in the pre-edge region. 
EELS spectra are generated from an integrated block of $3\times3$ pixels (spectra) in each case. 
EELS maps are generated by extracting the signal from a 3~eV energy window on the EELS spectrum. 
Specifically, energy windows of 283~eV to 286~eV, 290~eV to 293~eV, and 539~eV to 542~eV are used to generate maps for amorphous carbon, diamond, and oxygen, respectively. 
These are qualitative maps only, and careful deconvolution of the edge shapes and corrections for multiple scattering are not performed. 
As noted previously, we left the lamella slightly thicker due to the mechanical structure of the nanochannel; therefore, this EELS data is not optimal or usable for further quantitative analysis. 
The amorphous carbon can be quite easily isolated from the diamond carbon, as it exhibits a distinctive pre-peak at lower energies, which is absent in diamond.

The POCAL Python library \cite{Fontanot2023} is used for computing the simulated reflectance values. 
The real and complex refractive index values of the PCD film used in these computations are the same as those used in the Optical NanoGauge system, as measured by Hamamatsu Photonics.
Those used for water are taken from the literature \cite{Daimon2007}, and the real and complex refractive index values used for the Corning Lotus NXT glass substrate are fixed at 1.53 and 0, respectively.

\subsection*{Data availability}
Data and scripts for generating plots:\\
\href{https://github.com/StoffelJanssens/Nanochannels_2026}{https://github.com/StoffelJanssens/Nanochannels\_2026}

\subsection*{Supplementary information}
Supplementary material can be found in ancillary file:\\
\texttt{Supplementary\_information.pdf}

\subsubsection*{Acknowledgments}
This project is supported by OIST, with subsidy funding from the Cabinet Office, Government of Japan.
The authors thank the Engineering Section at OIST for the assistance and technical guidance.

\subsection*{Conflict of interests}
The authors declare that they have no known competing financial interests or personal relationships that could have appeared to influence the work reported in this paper.

\subsection*{Author contributions}
{\bf Stoffel D.\ Janssens:} Conceptualization, Methodology, Software, Validation, Formal analysis, Investigation, Data curation, Writing -- Original draft, Visualization. {\bf Meissha Ayu Ardini:} Investigation, Formal analysis, Writing -- Review and editing. {\bf David V\'azquez-Cort\'es:} Methodology, Investigation, Writing -- Review and editing. {\bf Cathal Cassidy:} Methodology, Validation, Formal analysis, Investigation, Data curation, Writing -- Review and editing, Visualization. {\bf Eliot Fried:} Writing -- Review and editing, Validation, Formal analysis, Project administration, Resources.

\makeatletter
\renewcommand{\@seccntformat}[1]{}
\makeatother

\renewcommand{\thesection}{}

\newpage

\subsection*{Table of Contents}
\begin{figure}[h!]
\centering
\includegraphics[scale=1]{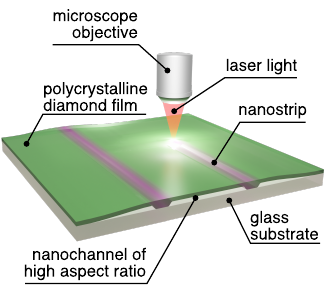}
\end{figure}
This work demonstrates the direct laser writing of nanochannels with aspect ratios exceeding fifty in the width-to-height ratio.
The channels have rectangular cross-sections, are optically accessible, can be filled with water, and are fabricated between diamond films and glass substrates.
The shape, microstructure, reflectance, and stability of the channels are investigated in detail.

\end{document}